\documentclass[a4,aps,amsmath,floatfix,nofootinbib,10pt]{revtex4} 
\usepackage{graphicx} \usepackage{color} \usepackage{enumerate} 
\newcommand{\be}{\begin{equation}} \newcommand{\ee}{\end{equation}} 
\newcommand{\ba}{\begin{array}} \newcommand{\ea}{\end{array}} 
\newcommand{\bea}{\begin{eqnarray}} \newcommand{\eea}{\end{eqnarray}} 
\newcommand{\bdm}{\begin{displaymath}} 
\newcommand{\edm}{\end{displaymath}} 
 
\newcommand{\sgn}{\operatorname{sgn}} 
 \begin{document}

\title{Hysteresis in the Ising model with Glauber dynamics.}

\author{Prabodh Shukla}

\affiliation{North Eastern Hill University \\ Shillong-793 022, India}

\begin{abstract}

We use Glauber dynamics to study time and temperature dependence of 
hysteresis in the pure (without quenched disorder) Ising model on 
cubic, square, honeycomb lattices as well as random graphs. Results are 
discussed in the context of more extensive studies of hysteresis in the 
random field Ising model.

\end{abstract}

\maketitle

\section{introduction}

The purpose of this note is to report work on temperature-driven 
hysteresis ~\cite{bertotti} in pure (i.e. without quenched disorder) 
Ising model ~\cite{ising,onsager} and compare it with more extensively 
studied case of disorder-driven hysteresis in the zero-temperature 
($T=0$) random field Ising model (ZTRFIM)~\cite{sethna1, maritan, 
sethna2, dhar, sethna3, perez, sethna4, xavier, rosinberg, liu, 
spasojevic, balog, shukla}. Hysteresis in ZTRFIM has been studied 
largely, if not entirely, in the limit of driving frequency $\omega \to 
0$. Normally hysteresis should vanish as $\omega \to 0$, but it 
survives because the limit $T \to 0$ is taken before the limit $\omega 
\to 0$. This is implemented by using $T=0$ Glauber dynamics 
~\cite{glauber} to update spins, and holding the applied field $h$ 
constant during updates. Thus one starts with all spins down in a 
sufficiently large and negative $h$, increases $h$ slowly till one spin 
flips up and causes a connected cluster of spins surrounding it to flip 
up in an avalanche. When the avalanche stops, $h$ is increased again 
until another avalanche occurs. The entire hysteresis loop is 
determined in this way by changing $h$ minimally between avalanches but 
keeping it fixed during avalanches.The dynamics of ferromagnetic ZTRFIM 
is Abelian. The order in which unstable spins are flipped does not 
matter. The stable configuration at $h$ is the same whether we reach it 
through a series of avalanches as described above or in one big 
avalanche starting from an initial state with all spins down.

A common choice for the random field distribution is a Gaussian with 
average zero and standard deviation $\sigma$. On simple cubic and 
several other lattices, there exists a critical value of 
$\sigma=\sigma_c$ that marks a phase transition in the response of the 
system to the applied field. For $\sigma > \sigma_c$ the magnetization 
$m(h)$ is smooth function of $h$, but for $\sigma < \sigma_c$ it 
acquires discontinuities at $h=\pm h_c$. The discontinuities reduce in 
size with increasing $\sigma$ and vanish continuously as $\sigma \to 
\sigma_c$. Extensive numerical and analytic work has established scale 
invariance and universality of phenomena in the vicinity of the 
non-equilibrium critical points $\{\pm h_c,\sigma_c\}$ in close parallel 
to the equilibrium critical behavior seen in the pure Ising model near 
the critical temperature $T_c$ ~\cite{wilson}. Indeed the parameter 
$\sigma$ in the ZTRFIM plays a role analogous to temperature $T$ in the 
pure Ising model. Although this similarity is well known, to the best 
of our knowledge, it has not been tested directly by simulating the 
$\omega$-dependent hysteresis loops in pure Ising model on a regular 
lattice.

We consider the kinetic Ising model on a cubic lattice characterized by 
the Hamiltonian,

\bdm H=-J\sum_{i,j}s_i s_j-h\sum_i s_i \edm

Here $J$ is ferromagnetic interaction between nearest neighbor Ising 
spins $\{s_i=\pm1\}$ situated on sites $\{i=1,2,\ldots,N\}$, and $h$ is 
a uniform applied field measured in units of $J$. We assume the system 
is in contact with a heat reservoir at temperature $T$. The Glauber 
prescription for updating a configuration $\{s_i\}$ is: (i) choose a 
site at random, say site $i$, (ii) calculate the local energy at site 
$i$, $e_i= -J \sum_{j \ne i} s_j -h$, (iii) flip $s_i$ to $-s_i$ with 
probability $1/(1+exp(-2Ke_i))$, where $K=J/k_BT$ and $k_B$ is the 
Boltzmann constant, (iv) repeat the above procedure $N$ times to 
complete one Monte Carlo cycle (unit of time), and (v) continue for $t$ 
Monte Carlo cycles. The above dynamics has two important properties; 
detailed balance and ergodicity. These properties combine to thermalize 
the system with increasing $t$. The dynamics of a fully thermalized 
system in the limit $t \to \infty$ generates configurations which are 
distributed according to their respective Boltzmann weights and are 
therefore uncorrelated with the initial configuration $\{s_i\}$ at 
$t=0$. The time average of a thermodynamic quantity over a sufficiently 
large number of such configurations should approximate to the 
corresponding equilibrium value obtained from the partition function of 
the system. An exact calculation of partition function is generally not 
feasible, so Glauber's or some other similar dynamics is the only 
practical way to explore equilibrium behavior of a system. However, 
predicting equilibrium behavior from dynamics has its own difficulties. 
Numerical studies are necessarily restricted to finite $t$ while 
equilibrium properties correspond to $t \to \infty$. It is not easy to 
decide what value of $t$ is adequate to extrapolate the results to 
equilibrium behavior. The answer depends on the temperature of the 
system and whether it is above or below the critical temperature. Our 
interest in the present paper is primarily in hysteresis which is a 
non-equilibrium phenomenon seen at finite $t$ only. We will examine how 
hysteresis decreases with increasing $t$ and if this behavior is 
consistent with the equilibrium behavior of the system reported in the 
literature. The equilibrium magnetization per site $m(h) = \sum_is_i/N$ 
depends on $K$. There is a critical value $K_c= 0.22165435(45)$ on a 
simple cubic lattice which marks the onset of spontaneous symmetry 
breaking in the system ~\cite{binder, sonsin, preis,gupta,haggkvist}. 
In the limit $t \to \infty$, at $h=0$, $m(h=0) \to 0$ if $K < K_c$, but 
$m(h=0) \to \pm m^*(K)$ with equal probability if $K \ge K_c$, where 
$|m^*(K)|$ increases continuously from zero to unity as $K$ increases 
from $K=K_c$ to $K=\infty$.

Hysteresis is generally characterized by a system's response to a 
cyclic field, but we may also consider it as a measure of system's 
memory of its initial state for $t < \infty$. It has been studied in 
several ways depending upon how the cyclic field is ramped up and 
down~\cite{rao, samoza, thomas, zheng}. We choose a method close in 
spirit to the one used in the ZTRFIM. We fix $K$ and $h$, and evolve 
two initial states $\{s_i=-1\}$ and $\{s_i=1\}$ separately for time 
$t$. Initially, the two states have magnetization per site equal to 
$-1$ and $+1$ respectively. Let the corresponding values at time $t$ be 
$m_-(K,h,t)$ and $m_+(K,h,t)$. Our simulations show that $m_+(K,h,t) > 
m_-(K,h,t)$ and the difference $m_+(K,h,t) - m_-(K,h,t)$ decreases with 
increasing $t$. This is to be expected from the properties of dynamics 
mentioned earlier. As the system approaches thermalization, the output 
configurations of the dynamics become uncorrelated with the initial 
configurations. Hysteresis becomes negligible at very large negative or 
positive values of $h$ $(|h| >> J)$ even for relatively small $t$ 
because the probability that a spin remains aligned opposite to a very 
large field is exponentially small. Thus, if $|h| >> J$, 
$m_-(K,h,t)\approx m_+(K,h,t)\approx \sgn h$.

\section{Numerical Results}

In our simulations, we choose a large range of $h$ around $h=0$, 
$[-H_0,H_0]$, such that induced magnetizations at $\pm H_0$ are nearly 
$\pm 1$. We divide the interval $[-H_0,H_0]$ into $n$ equal parts of 
width $\delta h=2H_0/n$, and calculate $m_-(K,h,t)$ and $m_+(K,h,t)$ at 
each increment $h_i=-H_0+i \times \delta h; i=0,\ldots,n$. We choose 
$n$ to be reasonably large so that the locus of data points on the 
graph indicates the shape of a continuous curve in the limit $n \to 
\infty$. We may call this curve the hysteresis loop at characteristic 
time period $t$ because each point on the curve has evolved for a time 
$t$ under the relaxation dynamics. The results are shown in Fig.1 for 
three values of $K$ in the vicinity of $K_c$ and two values of $t$ for 
each $K$. An interesting feature of Fig.1 is that the curves for 
$m_-(K,h,t)$ and $m_+(K,h,t)$ move towards each other as $t$ increases 
and look qualitatively similar to the hysteresis loops produced by a 
driving field of the form $h(t)=-H_0 \cos{\omega t}$, or a field that 
is ramped up and down in the form $h(t)=-H_0 + \omega t$ and $h(t)=H_0 
- \omega t$ respectively. Here $\omega=\Delta h / \Delta t$, and 
relaxation dynamics is applied to a configuration $\{s_i\}$ for a time 
$\Delta t$ at $h$ and the output is used as input at $h+ \omega \Delta 
t$. The fact that different methods produce similar hysteresis loops 
suggests that the generic $s$-shape of hysteresis loops comes from the 
probability distribution used in the relaxation dynamics rather than 
the form of the driving field. Notwithstanding the similarity of 
hysteresis loops, detailed behavior does depend on how the applied 
field is varied. A detail that interests us particularly is the 
variation of coercive field $H_c$ with $t$ for a given $K$. Fig.1 
indicates the trend that $H_c$ moves towards $h=0$ with increasing $t$. 
More detailed study requires monitoring the system at each value of $t$ 
and much smaller increments $\delta h$ in the applied field $h$ in the 
vicinity of $H_c$. This increases the computation time enormously and a 
compromise has to be made with respect to the size of $\delta h$. We 
have used $\delta h=0.001$ for cubic lattice and $\delta h=0.01$ for 
other lattices. This means that coercive fields in the range $0 < H_c 
\le \delta h$ will be clubbed at $H_c=\delta h$ in the respective 
cases. We will return to this point when discussing Fig.2, Fig.3, and 
Fig.4.

\begin{figure}[ht] 
\includegraphics[width=0.75\textwidth,angle=0]{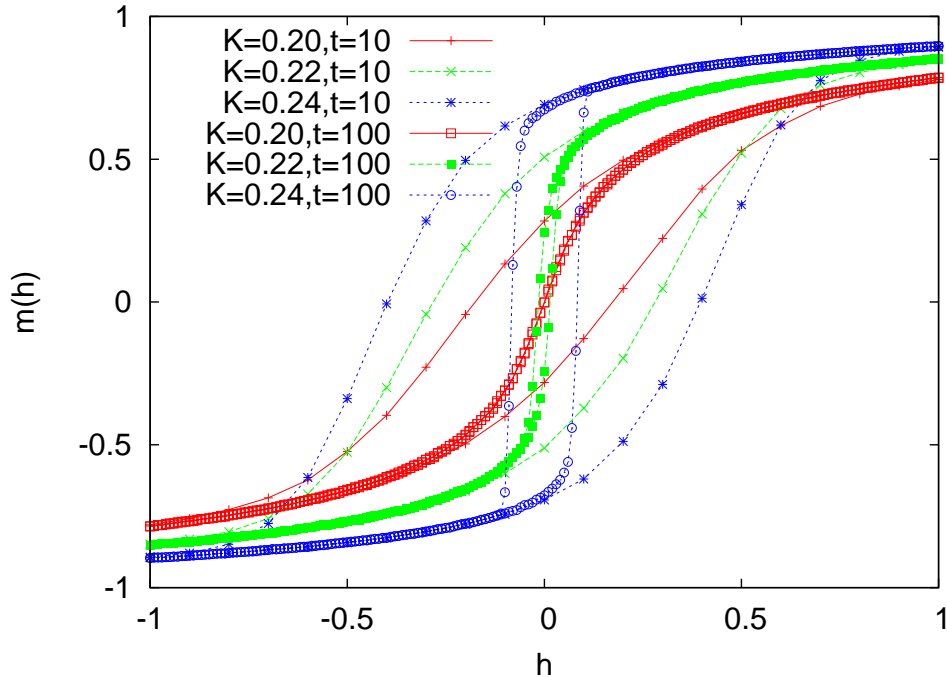} \caption{ 
Six hysteresis loops ($K=0.20, 0.22, 0.24;t=10, 100$ for each $K$) for 
pure Ising model on a $50^3$ cubic lattice near $K_c \approx 0.221654$. 
The figure suggests that hysteresis vanishes as $t \to \infty$ but it 
vanishes differently for $K < K_c$ than for $K>K_c$. For $K=0.20, 
t=100$, hysteresis has already vanished on the scale of the figure and 
$m(h)$ is continuous at $h=0$. For $K=0.24$, hysteresis at $t=100$ has 
reduced from its value at $t=10$ and $m(h)$ curves have come closer to 
vertical near the coercive field $H_c$. In this case, hysteresis is 
expected to vanish as $t \to \infty$ accompanied by a discontinuity in 
$m(h)$ at $h=0$.} \label{fig1} \end{figure}

Fig.1 shows hysteresis loops for $N=50^3$, $K=0.20, 0.22$, $0.24$, 
$t=10$, and $100$. It reveals two features of general validity. 
Firstly, hysteresis decreases with increasing $t$. For each $K$, the 
loop shrinks as we go from $t=10$ to $t=100$. The shrinking is faster 
if $K<K_c$ and the rate of shrinking increases with increasing $K_c-K$. 
At $t=100$ the loop is hardly visible for $K=0.22$ and not at all for 
$K=0.20$ on the scale of the figure. Secondly, the shape of loop 
changes with increasing $t$. It changes differently for $K<K_c$ than 
for $K>K_c$. If $K<K_c$, the loop tends to become narrower and 
elongated along the $x$-axis with increasing $t$. Eventually the loop 
collapses into a single continuous curve that passes through the origin 
$h=0$. If $K>K_c$, the loop becomes narrower and elongated along the 
$y$-axis with increasing $t$. In this case the middle portion of the 
$s$-shaped magnetization curve $m(h)$ becomes nearly vertical at the 
coercive field $H_c$ which moves very slowly towards $h=0$ with 
increasing $t$. In simulations, fluctuations make it rather difficult 
to distinguish between a continuous but steep change in $m(h)$ from a 
discontinuity in $m(h)$. We have checked this point carefully and 
conclude there is no discontinuity in $m(h)$ for finite $t$ up to the 
largest $t$ that we could test. In case there appeared to be a 
discontinuity in $m(h)$ at $H_c$, we re-examined the neighborhood of 
$H_c$ more closely by increasing the system size and decreasing the 
spacing $\delta h$ between neighboring $h$ values in the vicinity of 
$H_c$. This generated new data points inside the apparent discontinuity 
and indicated a sharply rising but continuous $m(h)$. We may also add 
that there is no theoretical reason to expect a true discontinuity in 
$m(h)$ at any finite $t$ for $T>0$. This suggests an approximate 
picture of the hysteresis loop as a flagpole with one dimensional flags 
at both ends but in opposite directions. The length of the flagpole 
increases with increasing $K-K_c$ and its width decreases with 
increasing $t$. As $t \to \infty$, we may expect the two halves of the 
loop to collapse on top of each other and the flagpole replaced by a 
discontinuity in $m(h)$ at $h=0$.

The above discussion suggests that the manner in which hysteresis 
decreases with increasing $t$ can reveal if the system is above or 
below its critical point. One can either monitor the rate at which the 
area of the hysteresis loop decreases, or alternately, how the coercive 
field decreases with increasing $t$. Fig.2 shows $H_c$ vs. $t$ on the 
lower half of the hysteresis loop for $N=50^3$, $K = 0.10, 0.20, 0.22, 
0.221654, 0.24, 1.00$, $1 \le t \le 2048$, and $0.001 \le H_c \le 4$. 
We have used logarithmic binning to reduce fluctuations in the data at 
large $t$ and drawn a line $H_c=1/t$ for comparison. It takes a good 
deal of computer time (several days) to generate the data shown in 
Fig.2 and it is the best we can do within our resources. So we look 
closely for possible trends in the data even if these trends are not as 
clear as we would desire. We see that four graphs corresponding to $K 
\le K_c$ are closer to each other and different from two graphs for $K 
> K_c$. If $K << K_c$, thermal fluctuations are very 
large and consequently the equilibrium correlation length is very 
short. Therefore the system relaxes to a thermalized state in a short 
time. As $K \to K_c$, the correlation length increases and so does the 
time to thermalization. Eventually at $K=K_c$ the correlation length 
and the time to thermalization diverge algebraically. At a given $t$, 
we may consider the magnitude of coercive field $H_c$ as a measure of 
the distance from equilibrium. Therefore we may expect $H_c$ to vary 
with $t$ as a power law at $K=K_c$. Our data indicates $H_c \sim 
t^{-1.15}$ approximately over two decades of $t$. Graphs for $K < K_c$ 
appear to behave similarly after an initial transient period and before 
fluctuations blur the trend. In the limit $K \to 0$ and $t \to \infty$, 
the spins would tend to flip independently of each other. System of 
$N=50^3$ will have random fluctuation in $m(h)$ of the order of 
$N^{-1/2} \approx 0.003$ around the value $m(H_c)=0$. The coercive 
field required to reverse the magnetization will have similar 
fluctuations. This is the reason why $H_c$ for $K=0.10$ shows a plateau 
at $H_c \approx 0.003$ and $t > 512$. Similarly in the case of $K=0.22$ 
and $K=0.22165$, plateaus are seen at $H_c \approx 0.002$ at $t > 
1024$. A plateau in $H_c$ at large $t$ could be expected for $K=0.20$ 
as well but its absence is within expected fluctuations. Next we turn 
our attention to graphs for $K > K_c$.In this regime thermal 
fluctuations diminish and long range order develops through nucleation, 
growth, coarsening of domains, and magnetization reversal. These 
processes are exceedingly slow due to smallness of thermal excitations. 
A large fraction of Glauber moves fail to flip the spins thus retarding 
the evolution of the system. Consequently $H_c$ decreases more slowly 
with increasing $t$ if $K>K_c$ than it does if $K < K_c$. The 
equilibrium value of order parameter increases with increasing $K$. 
Thus a magnetization reversal curve from a metastable state to a stable 
state at $H_c$ takes a nearly vertical shape in the vicinity of $H_c$ 
if $K>>K_c$. It takes a long time for relaxation dynamics to reverse 
the magnetization. A droplet of up spins has to nucleate in a sea of 
down spins and slowly grow to the system size. This is a very slow 
process and becomes progressively slower as $K$ increases. In the range 
of $t=1$ to $2024$, $H_c$ drops from $3.03$ to $0.05$ if $K=0.24$ and 
$3.47$ to $1.83$ if $K=1.00$. There is no good evidence that $H_c$ 
decreases with $t$ as a power law, nor we know of any theoretical 
reason to expect so. However, we note for later discussion that this 
very slow decrease of $H_c$ with $t$ for $K > K_c$ is a signature of a 
discontinuity in $m(h)$ at $h=0$ in the limit $t \to \infty$. The point 
is that for $K > K_c$ equilibrium $m(h)$ vs. $h$ curve must have a 
discontinuity at $h=0$ on grounds of symmetry breaking but it is 
difficult to observe the sharpness of this discontinuity via hysteresis 
dynamics in the limit $t \to \infty$. To the best of our knowledge, the 
limit $t \to \infty$ for this purpose has not been accessed even with 
the best of computers and most efficient Monte Carlo codes particularly 
for $K >> K_c$. The relaxation of the system is just too slow to reach 
equilibrium on practical time scales for $K >> K_c$. Quite often the 
sharpness of a first order transition is replaced by hysteresis in 
simulations as well as laboratory experiments. Thus we have to be 
content with the signature indicated above for the interpretation of 
our results in this regime.

\begin{figure}[ht] 
\includegraphics[width=0.75\textwidth,angle=0]{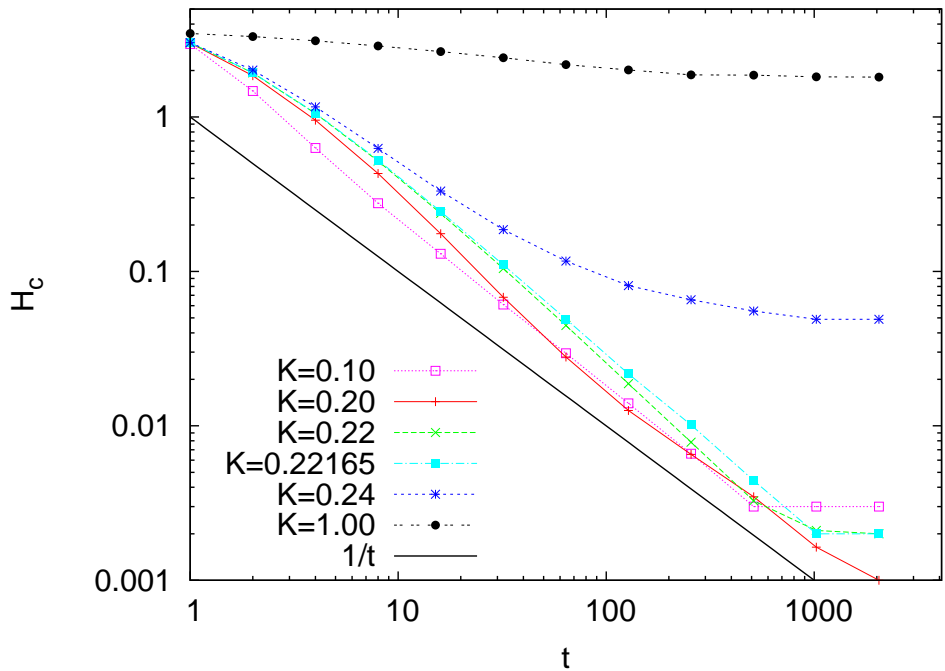} \caption { 
Coercive field $H_c$ on lower half of hysteresis loop after $t$ Monte 
Carlo cycles of Glauber dynamics on a $50^3$ cubic lattice starting 
with all spins down. A line $H_c=1/t$ is drawn for reference. If 
$K<K_c$, $H_c$ vanishes rapidly as $t \to \infty$. If $K > K_c$, $H_c$ 
decreases more and more slowly as $K$ increases. The reason is that 
$m(h)$ curve near $H_c$ tends to become vertical and it takes much 
longer time to reverse magnetization at $H_c$. It is a signature that 
$m(h)$ may acquire a discontinuity at $h=0$ in the limit $t \to \infty$ 
if $K > K_c$.} \label{fig2} \end{figure}

\begin{figure}[hb] 
\includegraphics[width=0.75\textwidth,angle=0]{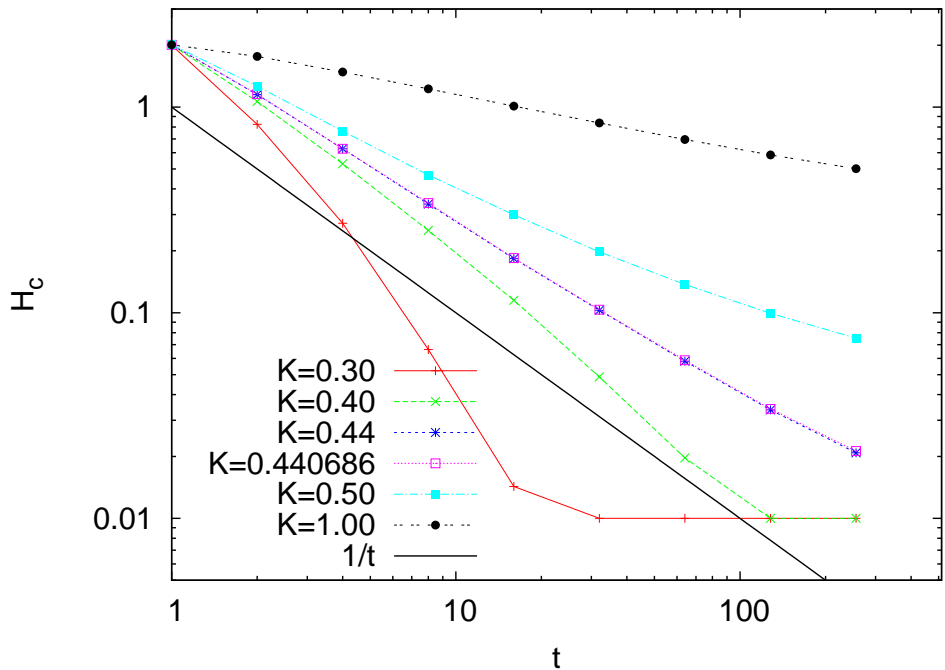} \caption 
{ Coercive field $H_c$ on a $500 \times 500$ square lattice for 
different values of $K$ and time periods $t$. The qualitative behavior 
on the square lattice is the same as on the cubic lattice. It indicates 
that $m(h)$ curve may acquire a discontinuity at $h=0$ in the limit $t 
\to \infty$ if $K > K_c$. A plateau is seen at $H_c=0.01$ because the 
system was monitored at increments of applied field equal to $0.01$.} 
\label{fig3} \end{figure}

We have also studied hysteresis and the variation of $H_c$ with $t$ on 
square and honeycomb lattices on the above lines. Ising models on these 
lattices are known to undergo equilibrium phase transitions at $K_c 
\approx 0.440686$ and $K_c=0.658479$ respectively. The exact values are 
given by $\sinh{2K_c}=1$ ~\cite{onsager} and $\cosh{2K_c}=2$ 
~\cite{wannier}. $H_c$ vs. $t$ graphs for the two cases are presented 
in Fig.3 and Fig.4 for lattices of size $500^2$, $1 \le t \le 512$, and 
$0.01 \le H_c \le 3$. Fig.3 shows the behavior on a square lattice for 
$K=0.30, 0.40, 0.44, 0.440686, 0.50, 1.00$. A line $H_c=1/t$ has been 
drawn for comparison. Graphs for $K=0.44$ and $K=0.440686$ are 
indistinguishable on the scale of figure and both vary approximately as 
$H_c \sim t^{-0.85}$. Graphs for $K=0.40$ and $K=0.30$ decay more 
rapidly but there is no discernible power law associated with the 
decrease. Two values of $K > K_c$, $K=0.50, 1.00$, indicate that $m(h)$ 
may hit a discontinuity at $h=0$ as $t \to \infty$ consistent with the 
known equilibrium phase transition in the system. We may add that an 
argument similar to the one used to explain the plateau in Fig.2 at 
$H_c \approx 0.003$ would lead us to expect plateaus in Fig.3 and Fig.4 
around $H_c \approx N^{-1/2}=0.002$ for large $t$ and $K << K_c$. But 
these are pre-empted by a larger step size $\delta h=0.01$ used in 
generating the data on square and honeycomb lattices. On these 
lattices, coercive fields in the range $0 < H_c \le 0.01$ are binned 
together resulting in a plateau at $H_c=0.01$. Because the plateau is 
an order of magnitude higher in comparison with Fig.2, time periods 
required to hit the plateau are smaller by an order of magnitude. This 
reduces the computer time without seriously compromising the general 
trends implicit in the data. In short, the behavior on the square 
lattice appears qualitatively similar to the behavior on the cubic 
lattice. Earlier we alluded to differences in hysteresis depending upon 
different forms of the driving field. Differences between sinusoidal 
and linear driving fields have been noted in the literature ~\cite{rao, 
samoza, thomas, zheng}. Fig.2 and Fig.3 provide another example. They 
indicate $H_c \sim t^{-1.15}$ on cubic, and $H_c \sim t^{-0.85}$ on 
square lattice at $K=K_c$. A subtle point to note is that the power-law 
fit on the square lattice is not quite as good as it is on the cubic 
lattice. A close look at Fig.3 shows that the critical curve turns up 
slightly at larger values of $t$. A possible explanation may be that 
$K_c$ on square lattice is obtained from the exact solution of the 
partition function while $K_c$ on cubic lattice is obtained from 
Glauber dynamics. It maybe that critical values of $K_c$ obtained from 
the two methods are somewhat different. This notwithstanding we can 
compare our power-laws with those for a field which is swept up from 
$-H_0$ to $H_0$ and back to $H_0$ in $t$ steps and at each step the 
previous output is used as the new input to dynamics. In this case, 
results for the area $A_c$ of the hysteresis loop at $K_c$ are 
available~\cite{zheng}. There is no exact relationship between $H_c$ 
and $A_c$ but they should be approximately proportional to each other 
in the limit $t \to \infty$. The reported results are $A_c \sim 
t^{-0.495}$ on cubic and $A_c \sim t^{-0.408}$ on square lattice which 
are significantly different from the power-laws observed in our version 
of the dynamics. We find hysteresis on honeycomb lattice to be 
qualitatively similar to that on cubic and square lattices. Fig.4 shows 
$H_c$ vs $t$ at $K_c=0.658479$, two values of $K<K_c$ and one value of 
$K>K_c$. The graphs for $K=K_c$ and $K>K_c$ show similar trend; both 
seem to be headed for an ordered state. This means that effective $K_c$ 
seen by dynamics is smaller than $K_c=0.658479$. The difference between 
$K_c$ and effective $K_c$ is larger on honeycomb lattice as compared 
with the same on square lattice.

\begin{figure}[ht] 
\includegraphics[width=0.75\textwidth,angle=0]{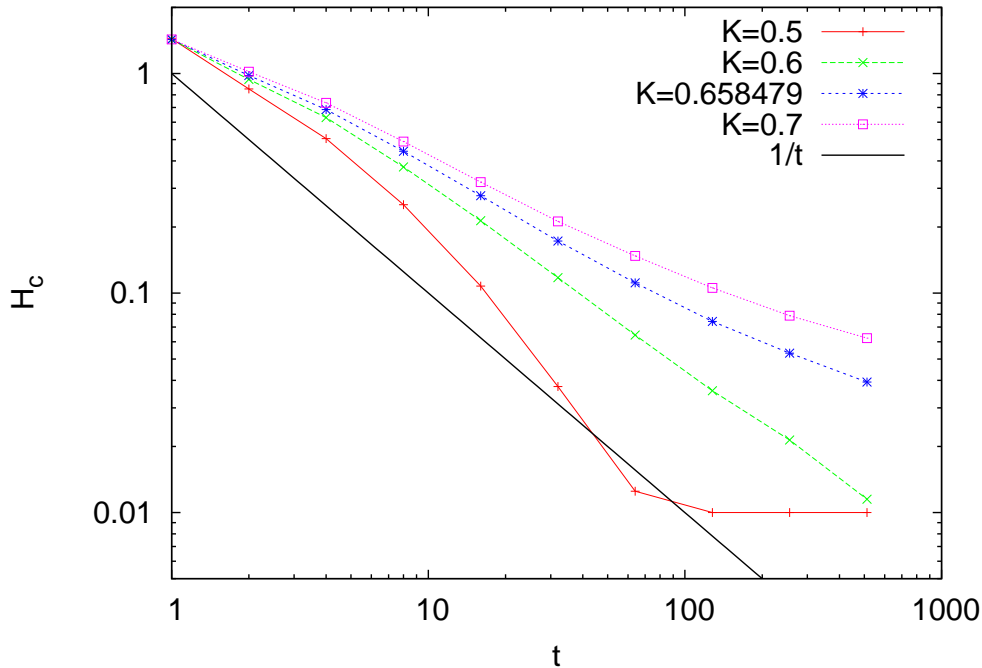} \caption{ 
Coercive field $H_c$ vs. $t$ on a $500\times500$ honeycomb lattice at 
different time periods $t$ for $K=0.5, 0.6, 0.658479, 0.7$. The figure 
indicates that effective transition point seen by the dynamics is 
smaller than $K_c=0.658479$.} \label{fig4} \end{figure}

In order to reconfirm and verify the trends indicated above, we have 
examined remanent magnetization $m_R$ on the lower half of hysteresis 
loop as a function of $t$; $m_R$ is the magnetization per site at $h=0$ 
starting from the initial state with all spins down. Time dependence of 
$m_R$ is relatively easy to monitor and it is a good indicator whether 
the system is evolving towards a disordered or an ordered state. If $K 
\le K_c$, we expect $m_R$ to decrease to zero with increasing $t$. This 
is born out by the results shown in Fig.5, Fig.6, Fig.7, and Fig.8. 
These figures show $m_R$ vs. $t$ on cubic, square, and honeycomb 
lattices, as well as on a random graph of coordination number $z=3$. 
Some noteworthy features are as follows. Irrespective of the lattice 
type, if $K << K_c$, correlation lengths are very short and $m_R$ 
approaches zero with increasing $t$ as expected. Consider Fig.5 for 
cubic lattice. At $K=K_c$, $m_R$ increases more slowly towards zero and 
somewhat surprisingly overshoots it at $t \approx 5000$. Such 
magnetization reversals are allowed within large critical fluctuations 
at $K=K_c$ but these should average out to zero eventually because the 
equilibrium value of the order parameter is zero. For $K > K_c$, we may 
expect $m_R$ to approach a finite value with increasing $t$ as indeed 
is seen in the Fig.5 for the cubic lattice. Fig.6 for square lattice 
shows similar behavior except for the case $K=K_c$ where there is some 
indication that $m_R$ may perhaps level off at a finite value for 
larger times. As mentioned earlier in reference to Fig.3, this may 
indicate that the effective $K_c$ for single-spin-flip Glauber dynamics 
may be somewhat smaller than $K_c=0.440686$. Fig.7 indicates similar 
behavior on honeycomb lattice. The effective $K_c$ recognized by 
dynamics on honeycomb lattice is apparently much smaller than 
$K_c=0.658479$. These results are consistent with the results shown in 
Fig.3 and Fig.4 and our interpretations of those figures. For reason to 
be discussed below we also examined $m_R$ on a random graph with 
coordination number $z=3$ which is a good representation of a Bethe 
lattice of the same connectivity. The exact $K_c$ on the Bethe lattice 
is given by the equation $\tanh{K_c}=(z-1)^{-1}$ ~\cite{rozikov}. The 
case $z=2$ corresponds to one dimensional Ising model which does not 
have a phase transition to an ordered state at any finite $K_c$. Thus, 
under the Glauber dynamics, remanent magnetization $m_R$ for $z=2$ 
should approach zero irrespective of $K$. We have verified that this is 
indeed the case. Of course, starting with all spins down, the time $t$ 
taken to thermalize increases with increasing $K$. For $z=3$, we have 
$K_c=0.54930615$ approximately. Remanence magnetization on 
corresponding random graph is shown in Fig.8. It indicates the 
occurrence of a phase transition in the system unlike the corresponding 
case in ZTRFIM. There is clear evidence that Glauber dynamics takes the 
system to an ordered state on a random graph of connectivity $z=3$ if 
$K>K_c$.

\begin{figure}[ht] 
\includegraphics[width=0.75\textwidth,angle=0]{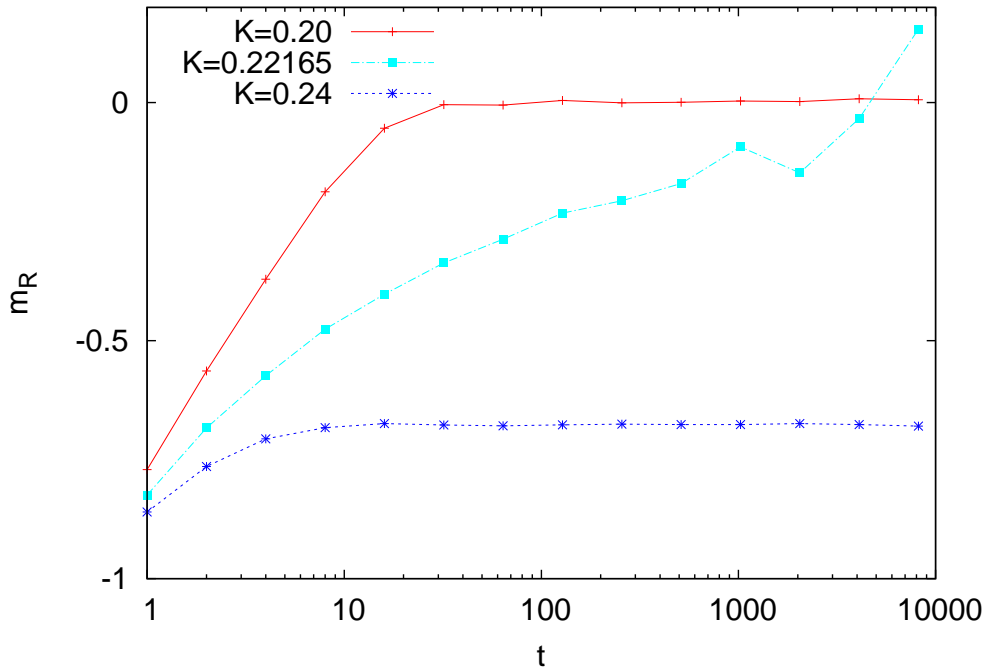} \caption 
{Remanent magnetization $m_R$ vs. $t$ on a $100^3$ cubic lattice on 
the lower half of the hysteresis loop for three representative values 
of $K$. As $t$ increases $m_R$ increases towards its equilibrium value; 
zero if $K \le K_c$, and a finite value if $K > K_c$. Magnetization 
reversal at $t \approx 5000$ for $K=K_c$ is a result of large critical 
fluctuations.} \label{fig5} \end{figure}

\begin{figure}[ht] 
\includegraphics[width=0.75\textwidth,angle=0]{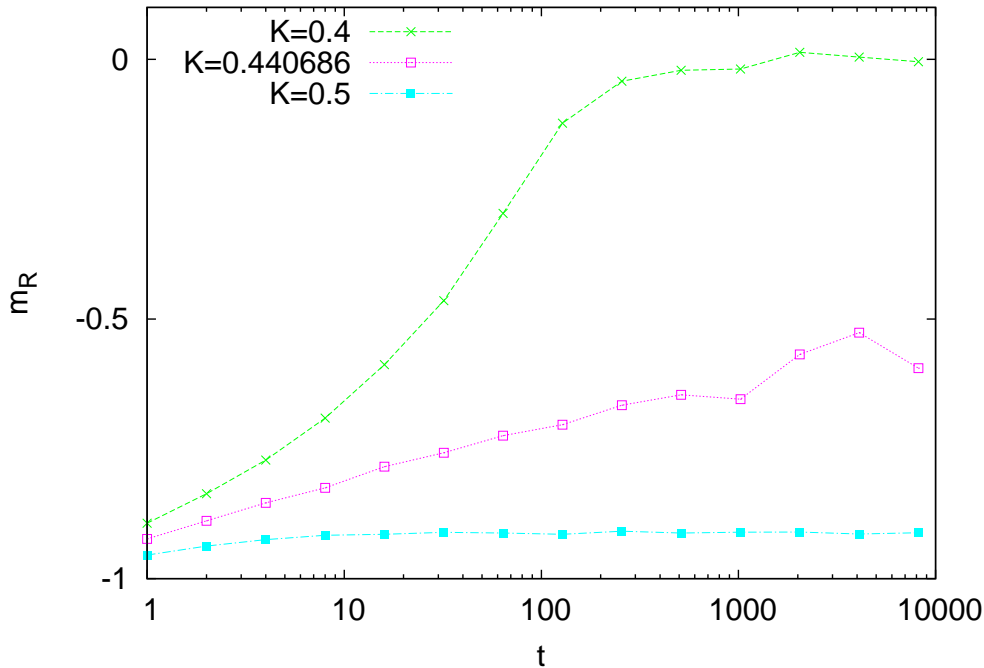} \caption 
{Remanence $m_R$ vs. $t$ on a $500 \times 500$ square lattice for three 
representative values of $K$. As $t$ increases $m_R$ is seen to 
approach its equilibrium value; zero if $K < K_c$ and a finite value if 
$K > K_c$. The limiting behavior for $K=K_c$ is not very clear. If 
$m_R$ eventually levels off at a finite value, it would indicate that 
the Onsager $K_c=0.440686$ is somewhat higher than appropriate $K_c$ 
for single-spin-flip Glauber dynamics. } \label{fig6} \end{figure}

\begin{figure}[ht] 
\includegraphics[width=0.75\textwidth,angle=0]{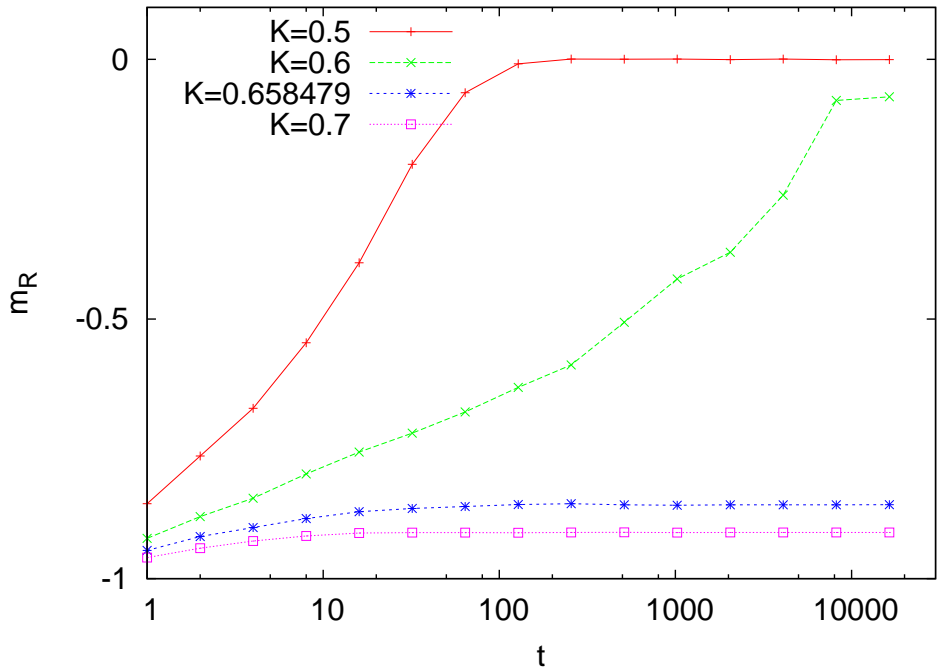} \caption 
{Remanence $m_R$ vs. $t$ on a $500 \times 500$ honeycomb lattice 
obtained from Glauber dynamics for four representative values of $K$. 
The critical value $K_c=0.658479$ is obtained from partition function; 
$m_R$ is expected to approach zero if $K \le K_c$, and a finite value 
if $K > K_c$. The plateau at $K_c$ suggests that the dynamical 
transition to an ordered state occurs at a higher temperature than 
predicted by the partition function. } \label{fig7} \end{figure}

\begin{figure}[ht] 
\includegraphics[width=0.75\textwidth,angle=0]{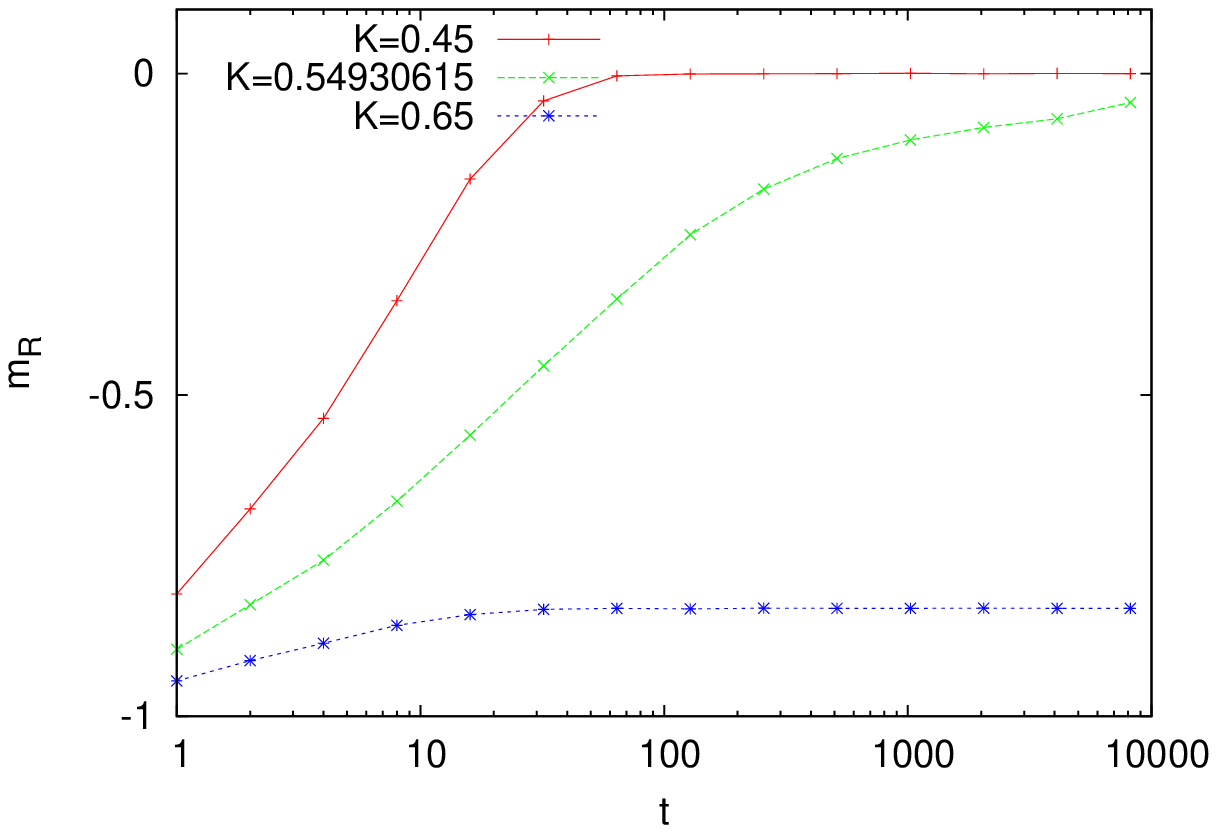} \caption 
{Remanence $m_R$ vs. $t$ on a random graph $(z=3, N=10^6)$ obtained 
from single-spin-flip Glauber dynamics. The critical value 
$K_c=0.54930615$ is obtained from partition function; $m_R$ is expected 
to approach zero if $K \le K_c$, and a finite value if $K > K_c$. The 
figure indicates a phase transition on a $z=3$ random graph unlike the 
case in ZTRFIM.} \label{fig8} \end{figure}

\newpage

\section{Discussion}

As mentioned in the Introduction, primary motivation for our study came 
from the curious similarity between nonequilibrium critical phenomena 
in ZTRFIM in the vicinity of $\sigma_c$ and equilibrium critical 
phenomena in pure Ising model in the vicinity of $K_c$. Does the 
similarity between $\sigma$ and $K$ hold only at the respective 
critical points or is it more extended? What would an extended 
similarity mean? We can not extract equilibrium behavior from 
hysteresis in ZTRFIM but we can explore equilibrium as well as 
nonequilibrium behavior of pure Ising model by supplementing it with 
finite temperature Glauber dynamics. The obvious thing is to compare 
phases on both sides of $\sigma_c$ with phases on both sides of $K_c$ 
and extend equilibrium studies in pure Ising model to nonequilibrium. 
The equilibrium $m(h)$ is a single valued anti-symmetric function of 
$h$, $m(h)=-m(-h)$, with a discontinuity at $h=0$ if $K > K_c$. There 
is no hysteresis in equilibrium i.e. the upper and lower halves of the 
hysteresis loop collapse on top of each other. On the other hand, 
$m(h)$ on each half of the hysteresis loop is discontinuous at the 
coercive field $H_c$ if $\sigma<\sigma_c$ and continuous for 
$\sigma>\sigma_c$. Could there be a relationship between the 
discontinuity in $m(h)$ for $\sigma < \sigma_c$ and the discontinuity 
in equilibrium $m(h)$? We may not expect such relationship at first 
because a system with quenched disorder is distinct from a system 
without quenched disorder. But critical behavior of both systems is 
similar. So it is possible that disorder whether quenched or thermal 
may produce qualitatively similar hysteresis away from the critical 
point as well.

Numerical results in the previous section suggest that disorder driven 
hysteresis is indeed qualitatively similar to temperature driven 
hysteresis at finite $t$ with a small difference. The discontinuity in 
hysteresis loops at the coercive field $H_c$ for $\sigma < \sigma_c$ is 
replaced by a continuous but steeply rising curve in the case of $K > 
K_c$. This is understandable. The discontinuities in $m(h)$ in ZTRFIM 
arise from quenched disorder as well as zero-temperature dynamics. The 
first provides local minima in the energy landscape and the second no 
escape from them except by changing the external field $h$. The field 
$h$ is assumed to vary infinitely slowly compared with the relaxation 
time of the system. Thus dynamics at each $h$ is allowed as much time 
as it takes to reach a locally stable state. In this framework it is 
possible for two arbitrarily close values of $h$ to have local minima 
with significantly different $m(h)$ i.e. a discontinuity in $m(h)$. A 
macroscopic discontinuity in $m(h)$ appears as an infinite avalanche in 
ZTRFIM. An infinite avalanche is also facilitated by zero-temperature 
dynamics because it does not allow a spin to flip back on the same half 
of the hysteresis loop. There are metastable states in pure Ising model 
as well but finite temperature Glauber dynamics enables the system to 
eventually evolve towards an equilibrium state. The state of the system 
at time $t$ in this case is determined by $K$. Therefore we may not 
expect two arbitrarily close values of $h$ to have very different 
magnetizations i.e. no discontinuity in $m(h)$ at any $h$ for finite 
$t$ and no infinite avalanches. Apart from the absence of a 
discontinuity in the hysteresis loop in the ordered phase, two phases 
separated by $K_c$ and $\sigma_c$ are similar. In the disordered phase 
($\sigma > \sigma_c, K < K_c$) correlation lengths are short and 
relaxation is fast while the opposite is true in the ordered phase. We 
may remark in passing that the dynamics of ZTRFIM is relatively fast 
even in the ordered phase because a spin once flipped does not flip 
back unless $h$ is reversed. For systems of same size, dynamics of 
magnetization reversal at $H_c$ for $\sigma < \sigma_c$ via an infinite 
avalanche in ZTRFIM is orders of magnitude faster than it is under 
Glauber dynamics of pure Ising model. Magnetization reversal at $H_c$ 
in the pure Ising model is so anomalously slow that we are often not 
able to complete it in simulations on practical time scales, especially 
for $K >> K_c$ but excluding $K=\infty$. The case $K=\infty$ is of 
course equivalent to ZTRFIM with $\sigma=0$.

ZTRFIM does not support a phase transition if the coordination number 
of the lattice is less than or equal to three i.e. $\sigma_c=0$ if 
$z\le3$. This result is based on an exact solution on Bethe lattice and 
simulations on periodic lattices with $z=3$ irrespective of the 
dimensionality of space in which they are embedded. This is puzzling at 
first sight. Usually Bethe lattices with $z>2$ behave similarly. The 
issue has been resolved for zero-temperature deterministic dynamics on 
Cayley trees which do not have closed loops. It has been argued that a 
minimal sprinkling of $z \ge 4$ sites on a spanning tree is required to 
sustain an infinite avalanche~\cite{shukla}. As the disorder is 
gradually increased to a critical value $\sigma_c$, the infinite 
avalanche vanishes at a nonequilibrium critical point. It is natural to 
ask if the absence of criticality on $z=3$ lattices persists under 
finite temperature Glauber dynamics of pure Ising model. With this in 
mind we studied hysteresis on honeycomb lattice and a random graph with 
connectivity $z=3$. In both cases the model undergoes an equilibrium 
phase transition. Simulations necessarily deal with finite $t$ and do 
not show a sharp transition on either lattice. But we find the 
qualitative behavior under finite temperature Glauber dynamics on $z=3$ 
lattices to be the same as on $z>3$ lattices. The system flows towards 
a disordered state for small $K$ and an ordered state for large $K$. It 
appears that the absence of criticality on $z=3$ lattices in ZTRFIM is 
perhaps an artifact of zero temperature dynamics and not intrinsic to 
lattice structure.

Our work indicates that critical value $K_c$ that separates two phases 
in the finite temperature dynamics is somewhat smaller than 
corresponding $K_c$ obtained from equilibrium statistical mechanics. It 
is not obvious why this should be so. The reason may lie in the 
limitations of one-spin flip dynamics. It is reasonable to assume that 
if two or more spins are allowed to flip jointly in one move the 
dynamics may take the system to a lower state of energy than is 
possible with one-spin flips. The broad features of phenomena including 
a phase transition seen with one-spin and two-spin flips may remain the 
same but overall energy scale may be pushed down somewhat in case of 
two-spin flips. Why this effect is larger on honeycomb lattice than it 
is on a square lattice or a random graph with $z=3$ requires further 
thought. This is a subtle reminder on the limitations of Monte Carlo 
methods. They average physical quantities on a smaller copy of system 
with has the same distribution of states as the full system. It is a 
bit like opinion polls which predict general elections. We may not 
expect an exact match between the two.

In summary, the work presented here complements extant studies of 
disorder driven hysteresis in ZTRFIM. Systems with extensive quenched 
disorder have thermodynamically large number of metastable states. The 
fact that disorder remains quenched implies that energy barriers 
between metastable states are much larger than thermal energy. On 
energy scale characterizing quenched disorder, it is reasonable to 
model hysteresis by ZTRFIM. Hysteresis loops in this model are obtained 
at $K=\infty$ and $t=\infty$. We have to bear in mind that none of 
these conditions are realized in a real experiment. Simulations reveal 
a discontinuity in $m(h)$ if $\sigma < \sigma_c$ and critical behavior 
at $\sigma =\sigma_c$. In simulations a discontinuity in $m(h)$ is 
often hard to distinguish from a very steep but continuous change in 
$m(h)$ but an exact solution of the model on a Bethe 
lattice~\cite{dhar} also supports the above scenario. We have shown 
that temperature driven hysteresis in a pure system is qualitatively 
similar to disorder driven hysteresis in ZTRFIM with minor differences. 
With increasing $t$, the remanent magnetization $m_R$ approaches zero 
if $K < K_c$ and a nonzero value if $K > K_c$. There is no 
discontinuity in $m(h)$ at the coercive field $H_c$ for $K>K_c$ 
although $m(h)$ curve does tend to become rather steep in the region 
around $H_c$ with increasing $t$. It would be satisfying to recover the 
expected equilibrium results by the dynamical route in the limit 
$t=\infty$ but this seems impossible on practical time scales due to 
ultra slow relaxation of the system. However this difficulty should not 
seriously compromise the applicability of this study to hysteresis 
experiments which are necessarily performed at a finite $K$ and finite 
$t$. Thus we hope results presented here will help in understanding a 
larger set of hysteresis experiments.

\end{document}